\documentclass[aps,prl,preprint,amssymb,groupedaddress,showpacs]{revtex4}
\usepackage{graphicx}
\usepackage{dcolumn}
\usepackage{amsmath}
\usepackage{amssymb}
\usepackage{subeqnarray}
\usepackage{bm}
\usepackage[usenames]{color}
\begin{document}

\title{Emergence of entanglement from a noisy environment: The case of polaritons}
\author{S. Portolan$^{1}$\footnote{Electronic address: stefano.portolan@epfl.ch},O. Di Stefano$^2$, S. Savasta$^2$, and V. Savona$^1$}
\affiliation{$^1$Institute of Theoretical Physics, Ecole
Polytechnique F\'{e}d\'{e}rale de Lausanne EPFL, CH-1015 Lausanne,
Switzerland} \affiliation{$^2$Dipartimento di Fisica della Materia e
Ingegneria Elettronica, Universit\`{a} di Messina Salita Sperone 31,
I-98166 Messina, Italy}

\begin{abstract}
We show theoretically that polariton pairs with a high degree of
polarization entanglement can be produced through parametric
scattering. We demonstrate that it can emerge in coincidence
experiments, even at low excitation densities where the dynamics is
dominated by incoherent photoluminesce. Our analysis is based on a
microscopic quantum statistical approach that treats coherent and
incoherent processes on an equal footing, thus allowing for a
quantitative assessment of the amount of entanglement under
realistic experimental conditions. This result puts forward the
robustness of pair correlations in solid-state devices, even when
noise dominates one-body correlations.
\end{abstract}


\maketitle

\newpage

The concept of \emph{entanglement} has played a crucial role in the
development of quantum physics. It can be described as the
correlation between distinct subsystems which cannot be reproduced
by any classical theory (i.e. \emph{quantum correlation}). It has
gained renewed interest mainly because of the crucial role that such
concept plays in quantum information/computation
\cite{Nielsen-Chuang}, as a precious resource enabling to perform
tasks that are either impossible or very inefficient in the
classical realm \cite{Amico et al RMP 2008 Plenio Virmani ent
meas2006}. Scalable solid-state devices will make use of local
electronic states to store quantum correlations \cite{DiVincenzo
rassegna}. Polaritons \cite{Rassegna Polaritons} on the other hand,
as hybrid states of electronic excitations and light, are the most
promising solution for generation and control of quantum
correlations over longer range \cite{Piermarocchi}. In particular,
thanks to the Coulomb interaction acting on the electronic part of
the polariton state, resonantly generated pump polaritons scatter
into pairs of {\em signal} and {\em idler} polaritons, in a way that
fulfills total energy and momentum conservation. The generated
polariton pairs are ideally in an entangled state \cite{Mandel,Ciuti
branch ent}. The peculiar energy-momentum dispersion of microcavity
polaritons has the advantage of allowing several configurations of
parametric scattering, that can be easily selected by setting the
frequency and angle of both the pump and the detected beams
\cite{Ciuti,Langbein PRB2004}.
In order to address entanglement in quantum systems \cite{Nature
CuCl, Edamatsu PRL}, the preferred experimental situations is the
few-particle regime in which the emitted particles can be detected
individually \cite{Portolan PRA2006}. In a real system, environment
always act as an uncontrollable and unavoidable continuous
perturbation producing decoherence and noise. Even if polariton
experiments are performed at temperature of few Kelvin \cite{Savasta
PRL2005}, polaritons created resonantly by the pump can scatter, by
emission or absorption of acoustic phonons, into other states,
acquiring random phase relations. These polaritons form an
incoherent background (i.e. noise), responsible of
\textit{pump-induced} photoluminescence (PL), which competes with
coherent photoemission generated by parametric scattering, as
evidenced by experiments \cite{Langbein PRB2004}. As a consequence,
noise represents a fundamental limitation, as it tends to lower the
degree of non-classical correlations or even completely wash it out
\cite{Zeilinger Nature2001,Savasta PRL2005}. Hence understanding the
impact of noise on quantum correlations in semiconductor devices,
where the electronic system cannot be easily isolated from its
environment, is crucial.

In this letter, we present a microscopic study of the influence of
time-dependent noise on the entanglement of polaritons generated in
parametric PL. Our treatment accounts for realistic features such as
detectors noise background, detection windows, dark-counting etc.,
needed \cite{Portolan PRB 2008a} in order to seek and limit all the
unwanted detrimental contributions. We show how a tomographic
reconstruction \cite{Qtomo Kwiat PRL-PRB 2001}, based on two-times
correlation functions, can provide a quantitative assessment of the
level of entanglement produced under realistic experimental
conditions. In particular, we give a ready-to-use realistic
experimental configuration able to measure the Entanglement of
Formation \cite{Bennett 1996 Wootters 1997 and 1998}, out of a
dominant time-dependent noise background, without any need for
post-processing \cite{Edamatsu PRL}.

\begin{figure}[!ht]
\begin{center}
\resizebox{8.0cm}{!}{\includegraphics{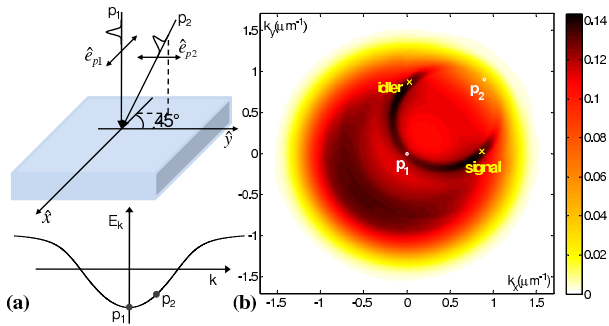}} \caption{(color
online)(a) sketch of the proposed excitation geometry and of the
lower polariton branch. The Gaussian pumps are linearly cross
polarized with zero time delay. The specific polarization
configuration with $\theta = 0$, where $\theta$ is the angle between
${\hat e}_{p1}$ and the $x$-axis on the $x-y$ plane, is depicted.
(b) The simulated spectrally integrated polariton population in
\textbf{k}-space. The parametric process builds up a circle passing
through the two pumps ($p_1$ and $p_2$), signal and idler polariton
states are represented by any two points on the circle connected by
a line passing through its center. For illustration the pair of
signal-idler polariton modes chosen for entanglement detection are
depicted as yellow crosses. The disc-shape contribution centered at
the origin is the incoherent population background produced by
phonon scattering.}\label{panel1}
\end{center}
\end{figure}

Third order nonlinear optical processes in quantum well excitons
(with spin $\sigma = \pm1$) can be described in terms of two
distinct scattering channels: one involving only excitons
(polaritons) with the same circular polarization (co-circular
channel); and the other (counter-circular channel) due to the
presence of both bound biexciton states and four-particle scattering
states of zero angular momentum ($J=0$) \cite{Sham PRL95}. Bound
biexciton-based entanglement generation schemes \cite{Nature
CuCl,Savasta SSC-Ishihara PRL2008}, producing entangled polaritons
with opposite spin, needs specific tunings for efficient generation,
and are expected to carry additional decoherence and noise due to
scattering of biexcitons. Moreover linearly polarized single pump
excitation (i.e. ${\hat e}_p = \cos \theta |x\rangle + \sin \theta
|y\rangle=2^{-1/2}(e^{i\, \theta} |-\rangle + e^{-i\, \theta}
|+\rangle)$) cannot avoid the additional presence of the co-circular
scattering channel which can lower polarization entanglement.
The experimental set-up that we propose is a two-pump scheme under
pulsed excitation, involving the lower polariton branch only. The
pumps ($p_1$ and $p_2$) are chosen with incidence angles below the
{\em magic angle} \cite{Langbein PRB2004} so that single-pump
parametric scattering is negligible. In this setup, mixed-pump
processes (signal at in-plane wave vector ${\bf k}$, idler at ${\bf
k}_i = {\bf k}_{1} + {\bf k}_{2} - {\bf k}$) are allowed, as shown
in Fig. \ref{panel1}. As signal-idler pair, we choose to study the
two energy-degenerate modes at ${\bf k}\simeq(k_{1x},k_{2y})$ and
${\bf k}_{i}\simeq(k_{2x},k_{1y})$. For all the numerical
simulations we will consider the sample investigated in Ref.
\cite{Langbein PRB2004}. We choose ${\bf k}_{1}=(0.,0.) \mu m^{-1}$,
and ${\bf k}_{2}=(0.9,0.9) \mu m^{-1}$.
%==========================================================================
In particular, we shall employ two pump beams linearly cross
polarized (then the angle $\theta$ will refer to the polarization of
one of the two beams, see Fig. (\ref{panel1})). This configuration
is such that the counter-circular scattering channel (both bound
biexciton and scattering states) is suppressed owing to destructive
interference, while co-circular polarized signal-idler beams are
generated. In the absence of the noisy environment, polariton pairs
would be cast in the pure triplet entangled state $|+,+\rangle -
\exp{(i 4 \theta)} |-,-\rangle$. On the opposite side, two-pump
excitation with counter-circularly pump beams (here not adopted) is
able to stop this channel giving rise to the entangled state
$|+,-\rangle - \exp{(i 4 \theta)} |-,+\rangle$.

%
%===================================================================================
The advantages of this configuration are manyfold. First,
detrimental processes for entanglement as the excitation induced
dephasing results to be largely suppressed \cite{biexciton
contribution}. Spurious coherent processes, e.g. Resonant Rayleigh
Scattering \cite{Langbein RRS}, are well separated in ${\bf
k}$-space from the signal and idler modes. In addition, this
configuration with signal and idler close to the origin in ${\bf
k}$-space makes negligible the longitudinal-transverse splitting of
polaritons \cite{Kavokin} (relevant at quite high in-plane wave
vectors).
%
%==================================================================
It is worth noticing that one could swap the roles of signal-idler
and pump pairs, resulting in a parametric setup with isoenergetic
pumps. However, such configuration would suffer from non-symmetric
dephasing and non-equilibrated signal-idler photon intensities, that
would lower the quantum correlation between the outgoing photons
\cite{Quattropani+Giacobino}.
%==========================================================================
%==========================================================================

Following Ref. \cite{Qtomo Kwiat PRL-PRB 2001} the tomographic
reconstruction of the two-polariton density matrix is equivalent, in
the $\sigma = \{+,-\}$ polarization basis, to the two-time
coincidence
\begin{equation}\label{rho gen} \hat{\rho}_{\sigma \tilde{\sigma}, \sigma' \tilde{\sigma}'}
= \frac{1}{\mathcal{N}} \hspace{-0.1cm} \int_{T_d} \hspace{-0.25cm}
dt_1 \hspace{-0.1cm} \int_{T_d}\hspace{-0.25cm} dt_2 \langle
\hat{P}^\dag_{{\bf k} \sigma}(t_1) \hat{P}^\dag_{{\bf k}_i
\tilde{\sigma}}(t_2) \hat{P}_{{\bf k}_i \tilde{\sigma}'}(t_2)
\hat{P}_{{\bf k} \sigma'}(t_1) \rangle\, ,
\end{equation}
where $\hat{P}^\dag_{{\bf k} \sigma}$ ($\hat{P}^\dag_{{\bf k}_i
\tilde{\sigma}}$) creates a signal polariton at ${\bf k}$ (an idler
polariton at ${\bf k}_i = {\bf k}_{1} + {\bf k}_{2} - {\bf k}$),
$\mathcal{N}$ is  a normalization constant and $T_d$ the detector
window. We choose a very wide time window $T_d=120\, \text{ps}$,
allowing feasible experiments with standard photodetectors. In order
to model the density matrix eq.\, (\ref{rho gen}), we employ the
dynamics controlled truncation scheme (DCTS), starting from the
electron-hole Hamiltonian including two-body Coulomb interaction and
radiation-matter coupling. In this approach nonlinear parametric
processes within a third order optical response are microscopically
calucated. The main environment channel is acoustic phonon
interaction via deformation potential coupling \cite{Portolan PRB
2008a, Piermarocchi PRB}. In order to account for coherent and
incoherent processes on an equal footing we use a DCTS-Langevin
approach \cite{Portolan PRB 2008a}, with noise sources given by
exciton-LA-phonon scattering and radiative decay (treated in the
Born-Markov approximation). For mixed-pump processes with arbitrary
polarization the Heisenberg-Langevin equations of motion read:

\begin{eqnarray}\label{sys 2pumps}
\frac{d}{dt} \hat{P}_{{\bf k} \sigma} = -i \tilde \omega_{\bf
k}\hat{P}_{{\bf k} \sigma} -i g_{|\sigma_1+ \sigma_2|}
\hat{P}^{\dag}_{{\bf k_i} \sigma_i}
\mathcal{P}_{{\bf k}_{1}\sigma_1}\mathcal{P}_{{\bf k}_{2}\sigma_2}+ \hat{\mathcal{F}}_{\hat{P}_{{\bf k} \sigma}}  \\
\frac{d}{dt} \hat{P}^\dag_{{\bf k_i} \sigma_i} =
 i \tilde\omega^*_{\bf k_i} \hat{P}^\dag_{{\bf k_i} \sigma_i} +
 i g_{|\sigma_1+ \sigma_2|} \hat{P}_{{\bf k}\sigma}
\mathcal{P}^{*}_{{\bf k}_{1} \sigma_1}\mathcal{P}^{*}_{{\bf
k}_{2}\sigma_2} + \hat{\mathcal{F}}_{\hat{P}^\dag_{{\bf k_i}
\sigma}}\, . \nonumber
\end{eqnarray}
In Eq. (\ref{sys 2pumps}) $\mathcal{P}_{{\bf k},\sigma}$ are the
projections onto the circular basis $\sigma$ of the coherent pump
polariton fields. The complex polariton dispersion $
\tilde{\omega}_{\bf k}$ includes the effects of relaxation and
pump-induced renormalization (shifts), $g$ is the nonlinear
interaction term driving the mixed parametric processes
\cite{Portolan PRB 2008a}; summation over the repeated polarization
indices $\sigma_1$ and $\sigma_2$ is assumed and the following
selection rule holds: $\sigma + \sigma_i = \sigma_1 + \sigma_2$. In
general, third order contributions due to Coulomb interaction
between excitons account for terms beyond mean-field, including an
effective reduction of the mean-field interaction and an excitation
induced dephasing. It has been shown that both effects depend on the
sum of the frequencies of the scattered polariton pairs
\cite{biexciton contribution}. For the frequency range here
exploited, the excitation induced dephasing is vanishingly small and
can be safely neglected on the lower polariton branch
\cite{biexciton contribution}. On the contrary, the matrix elements
of the ($J=0$) counter-circular scattering channel is lower (about
1/3) than that for the co-polarized scattering channel, but
certainly not negligible \cite{biexciton contribution}. However, in
the pump polarization scheme that we propose, performing the
polarization sum in Eq.\ (\ref{sys 2pumps}), it is easy to see that
the counter-circular channel cancels out, as already pointed out.
This feature is unique to the present scheme, while all previously
adopted pump configurations suffer from the presence of both singlet
and triplet channels. Equation (\ref{sys 2pumps}) is a system of two
coupled equations for polariton operators, acting onto the global
system and environment state space, thanks to the two additive noise
sources $\hat{\mathcal{F}}_{\hat{P}_{{\bf k} \sigma}}$,
$\hat{\mathcal{F}}_{\hat{P}^\dag_{{\bf k_i} \sigma}}$ \cite{Portolan
PRB 2008a,Lax,notaFDT}
%
%
%
%===================================================================================
\begin{figure}[!ht]
\begin{center}
\resizebox{8.0cm}{!}{\includegraphics{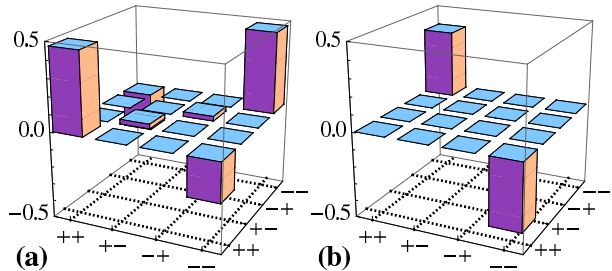}}\caption{(color
online) a(b) Real (Imaginary) part of the density matrix in the
tomographic reconstruction according to eq. (\ref{rho gen}) in the
linearly cross polarized pump configuration with $\theta = \pi / 6$.
This case results in $E(\hat{\rho}) \simeq 0.7523$}\label{parte
reale}.
\end{center}
\end{figure}
%===================================================================================
%

Early experiments in semiconductor microcavities
\cite{Giacobino,Savasta PRL2005} provided promising, though
indirect, indications of polariton entanglement. In order to achieve
a conclusive evidence of entanglement one has to produce its
quantitative analysis and characterization, i.e. a \emph{measure} of
entanglement. Among the various measures proposed in the literature
we shall use the \textit{entanglement of formation} (EOF)
$E(\hat{\rho})$ \cite{Bennett 1996 Wootters 1997 and 1998} for which
an explicit formula as a function of the density matrix exists. It
has a direct operational meaning as the minimum amount of
information needed to \textit{form} the entangled state under
investigation out of uncorrelated ones \cite{EOF}. The complete
characterization of a quantum state requires the  knowledge of its
density matrix. For a quantum system composed of two two-level
particles, it can be reconstructed by {\em quantum state tomography}
\cite{Nature CuCl,Qtomo Kwiat PRL-PRB 2001}. It requires 16
two-photon coincidence measurements based on various polarization
configurations \cite{Qtomo Kwiat PRL-PRB 2001}. Exploiting the Wick
factorization \cite{Mandel Zinn} and the symmetries of the system,
the density matrix elements are built up on the signal and idler
occupation $ N_{s/i \pm} \doteq \langle \hat{P}^\dag_{s/i \pm}(\tau)
\hat{P}_{s/i \pm}(\tau) \rangle$ and on the two-time correlation
functions $\langle \hat{P}^\dag_{s+}(u) \hat{P}^\dag_{i+}(v) \rangle
$ and $\langle \hat{P}^\dag_{s-}(u) \hat{P}^\dag_{i-}(v) \rangle$.
We integrate the system of equations eq. (\ref{sys 2pumps}), coupled
with the underlying non-equilibrium equations for the noise
%($\hat{\mathcal{F}}_{\hat{P}_{{\bf k} \pm}}$,
%$\hat{\mathcal{F}}_{\hat{P}^\dag_{{\bf k_i} \pm}}$)
%
correlation
functions (through time-dependent fluctuation-dissipation relations) %\cite{Portolan PRB 2008a,Lax}
\cite{Portolan PRB 2008a,Lax}. For a generic polarization $\theta$,
the populations are independent on $\theta$, $N_{s/i +} = N_{s/i -}
\doteq N_{s/i}$, whereas correlations satisfy $\langle
\hat{P}^\dag_{s+}(u) \hat{P}^\dag_{i+}(v) \rangle = - e^{i\, 4
\theta} \langle \hat{P}^\dag_{s-}(u) \hat{P}^\dag_{i-}(v) \rangle$.
\begin{figure}[!ht]
\begin{center}
\resizebox{!}{5.0cm}{\includegraphics{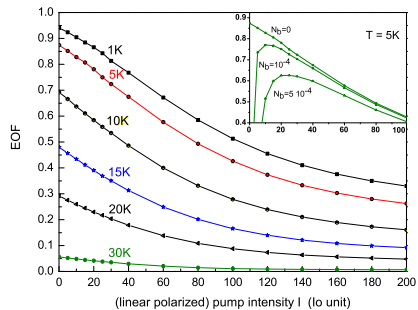}}\caption{(color
online) Dependence of the EOF on pumping intensity. The laser
intensity $I$ is measured in units of $I_0 = 21\, \text{photons} \,
\mu\text{m}^{-2} / \text{pulse}$ according to Ref.\, \cite{Langbein
PRB2004}.}\label{EOF_I}
\end{center}
\end{figure}
In Fig. \ref{parte reale} a tomographic reconstruction is shown. We
point out that different phase relations appearing in the
non-diagonal terms of a reconstructed density matrix are directly
related to the choice of pump linear polarization $\theta$ (see
caption of Fig.\, \ref{parte reale}). In the absence of
longitudinal-transverse splitting \cite{Kavokin} polariton
entanglement is independent of the angle of the linear pump
polarization $\theta$. As Fig. \ref{EOF_I} shows, there is a
non-negligible region of the parameter space where, even in a
realistic situation, high entanglement values are obtained. For
increasing pump, EOF decays towards zero. This is a known
consequence of the relative increase of signal and idler populations
\cite{Nature CuCl,Edamatsu PRL} -- dominating the diagonal elements
of $\hat{\rho}$ -- with respect to two-body correlations responsible
of the non-diagonal parts, which our microscopic calculation is able
to reproduce. We expect entanglement to be unaffected by both
intensity and phase fluctuations of the pump laser. The former are
negligible according to Fig. 3, while the latter only act on the
overall quantum phase of the signal-idler pair state.

Different entanglement measures generally result in quantitatively
different results for a given mixed state. However, they all provide
upper bounds for the distillable entanglement \cite{Amico et al RMP
2008 Plenio Virmani ent meas2006}, i.e. the rate at which mixed
states can be converted into the ``gold standard" singlet state.
Small EOF means that a heavily resource-demanding distillation
process is needed for any practical purpose. Fig. \ref{EOF_I} shows
how a relative small change in the lattice temperature has a
sizeable impact on entanglement. As an example, for the pump
intensity $I=15 I_0$, increasing the temperature from $T=1K$ to
$T=20K$ means to corrupt the state from $E(\hat{\rho}) \simeq 0.88$
to $E(\hat{\rho}) \simeq 0.24$, whose distillation is nearly three
times more demanding. For a fixed pump intensity, Fig. \ref{EOF_I}
shows that, above a finite temperature threshold, the EOF vanishes
independently of the pump intensity, i.e. the influence of the
environment is so strong that quantum correlation cannot be kept
anymore. In physical terms, at about $30 K$ the average phonon
energy becomes comparable to the signal-pump and idler-pump energy
differences, and the thermal production of signal-idler pairs is
activated.

\begin{figure}[!ht]
\begin{center}
\resizebox{!}{4.50cm}{\includegraphics{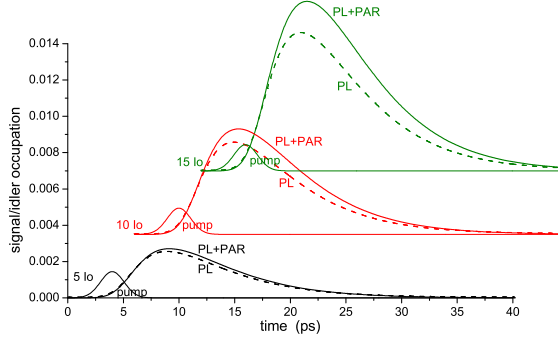}}\caption{(color
online) Signal/idler occupations (dashed PL only, full
PL+parametric) in time for three excitation densities ($I=5\ I_0$,
$I=10\ I_0$ and $I=15\ I_0$) for $T=5K$ are shown. The shape of the
pump pulse is depicted for reference.}\label{Ns}
\end{center}
\end{figure}

Semiconductors heterostructures are complex systems in which other
noise sources are expected. The simplest way to model this
additional noise is via the introduction of a constant, temperature-
and pump-independent, noise background $N_b$. This quantity also
accounts for the noise background characterizing the photodetection
system. In the inset of Fig. \ref{EOF_I}, the dependence of the EOF
on $N_b$ is highlighted. In our simulation, the quantity $N_b$
causes EOF to vanish in the limit of low pump intensity. As Fig.
\ref{Ns} shows, a value of $N_b$ of about $10^{-4}$ is realistic, as
suggested by experiments \cite{Savasta PRL2005} showing that it is
considerably smaller than the PL-noise studied here. From inspection
of the density matrix (in the $(++,+-,-+,--)$ circular-polarization
basis), for pump intensity ${I \rightarrow 0}$, if the correlation
dominates, we have as limiting case a triplet pure state
\begin{eqnarray}
  \rho \rightarrow \frac{1}{2} \left( \begin{array}{cccc}
1 & 0 & 0 & 1 \\
0 & 0 & 0 & 0 \\
0 & 0 & 0 & 0 \\
1 & 0 & 0 & 1 \end{array} \right) , \nonumber
\end{eqnarray}
on the other hand, if the population dominates, the state becomes
separable $ \rho \rightarrow \frac{1}{4} \mathbb{I}$. At leading
order in $I$, the two contributions are comparable, and EOF emerges
due to the normalization term $\mathcal{N}$. For nonzero $N_b$
instead, the leading term in the population is constant. This
explain the behaviour of EOF as ${I \rightarrow 0}$. Mathematically
speaking, for any finite value of $N_b$, we have a separable state
in the limit ${I \rightarrow 0}$. As seen in the inset of Fig.
\ref{EOF_I}, a finite $N_b$ affects only the range of very low pump
intensity $I$.
Figure 4 displays the time-resolved signal/idler occupations
calculated at $T = 5 K$ for three different excitation densities.
The dashed lines describe the PL contributions to the occupation.
The figure clearly shows that at low excitation density the detected
intensity in the signal/idler channels arises mainly from PL.
Nevertheless the obtained EOF for these intensities is very high,
contrarily to intuition, but in agreement with recent results
\cite{Edamatsu PRL}. This result puts forward the robustness of pair
correlations and entanglement that can be evidenced, even when noise
is the dominant contribution to one-body correlations.

In conclusion we have shown that microcavity polaritons can be cast
in an entangled state in a controlled way and we have given a
ready-to-use realistic experimental configuration able to measure
the EOF out of a dominant time-dependent noise background, without
any need for post-processing. We point out that the outcome of a
polariton parametric scattering is an {\em entangled state of an
electronic excitation of the semiconductor} -- the polariton pair --
contrarily to parametric downconversion in a nonlinear crystal
\cite{Mandel}, where only the outgoing photons are entangled. Here,
the emitted photons serve merely as a probe of the internal degree
of entanglement. Thanks to their photon component, polaritons can
sustain quantum correlations over mesoscopic distances {\em inside}
the semiconductor. This is why they bear a unique potential as a
controllable embedded mechanism to generate quantum information in a
device and transfer it to localized qubits (e.g. spin qubits) over
distances of microns \cite{Piermarocchi}.

\begin{acknowledgments}
S.P. acknowledges the support of NCCR Quantum Photonics (NCCR QP),
research instrument of the Swiss National Science Foundation (SNSF).
\end{acknowledgments}


\begin{thebibliography}{50}

\bibitem{Nielsen-Chuang} M. A.Nielsen, \& I. L. Chuang, {\em Quantum Computation and
Quantum Information} (Cambridge University Press, Cambridge, 2000).

\bibitem{Amico et al RMP 2008 Plenio Virmani ent
meas2006} L. Amico, \emph{et. al.}, Rev. Mod. Phys. {\bf 80}, 517
(2008); M. B. Plenio, \& S. Virmani, Quant. Inf. Comp. {\bf 7}, 1
(2007). % ; arXiv:quant-ph/0504163v3.

\bibitem{DiVincenzo rassegna} D. P. DiVincenzo, Science {\bf 270}, 255
(1995). %; D. P. DiVincenzo, Fortschr. Phys. {\bf 48}, 771-783
(2000).

\bibitem{Rassegna Polaritons} V. Savona, \emph{et. al.} Phase
Transitions {\bf 86}, 169-279 (1999); A. Kavokin, \& G. Malpuech,
Cavity Polaritons. Elsevier, Amsterdam (2003); G. Khitrova,
\emph{et. al.} Rev. Mod. Phys. {\bf 71}, 1591 (1999).

\bibitem{Piermarocchi} G. F. Quinteiro, \emph{et. al.}, Phys. Rev. Lett. {\bf 97}, 097401 (2006).

\bibitem{Mandel} L. Mandel, Rev. Mod. Phys. {\bf 71}, S274 (1999).

\bibitem{Ciuti branch ent} C. Ciuti, Phys. Rev. B {\bf 69}, 245304 (2004).

\bibitem{Ciuti} C. Ciuti, Phys. Rev. B {\bf 69}, 245304 (2004)

\bibitem{Langbein PRB2004} W. Langbein, Phys. Rev. B {\bf 70}, 205301
(2004).

%\bibitem{Sham PRL95} Th. \"{O}streich, \emph{et. al.}, Phys.
%Rev. Lett. {\bf 74}, 4698 (1995), Phys. Rev. B {\bf 58}, 12920
%(1998).

\bibitem{Nature CuCl} K. Edamatsu, \emph{et. al.}, Nature {\bf 431},
167-170 (2004).

\bibitem{Edamatsu PRL} G. Oohata, \emph{et. al.}, Phs. Rev. Lett.
{\bf 98}, 140503 (2007).

\bibitem{Portolan PRA2006} S. Portolan, \emph{et. al.}, Phys. Rev. A {\bf 73},
020101(R) (2006); R. J. Glauber, Phys. Rev. {\bf 130}, 2529 (1963).

\bibitem{Savasta PRL2005} S. Savasta, \emph{et. al.}, Phys. Rev. Lett. {\bf 94}, 246401 (2005).

\bibitem{Zeilinger Nature2001} J. W. Pan, \emph{et. al.}, Nature {\bf
410}, 1067-1070 (2001).

\bibitem{Portolan PRB 2008a} S. Portolan, \emph{et. al.}, Phys. Rev. B {\bf 77}, 035433
(2008); S. Portolan, \emph{et. al.}, Phys. Rev. B {\bf 77}, 195305
(2008).


\bibitem{Qtomo Kwiat PRL-PRB 2001} A. G. White, \emph{et. al.}, Phys. Rev. Lett. {\bf 83}, 3103 (1999).
%; D. F. V. James, \emph{et. al.}, Phys. Rev. B {\bf 64}, 052312 (2001).

\bibitem{Bennett 1996 Wootters 1997 and 1998} C. H. Bennett, \emph{et. al.}, Phys. Rev. A {\bf 54}, 3824
(1996)%; S. Hill, and W. K. Wootters, Phys. Rev. Lett. {\bf 78}, 5022
%(1997)
; W. K. Wootters, Phys. Rev. Lett. {\bf 80}, 2245 (1998).


\bibitem{Sham PRL95} Th. \"{O}streich, \emph{et. al.}, Phys.
Rev. Lett. {\bf 74}, 4698 (1995).%, Phys. Rev. B {\bf 58}, 12920
%(1998).


\bibitem{Savasta SSC-Ishihara PRL2008} S. Savasta, \emph{et. al.}, Solid State Communication, {\bf 111} 495
(1999); H. Oka, H. \& H. Ishihara, Phys. Rev. Lett. {\bf 100},
170505 (2008).



\bibitem{biexciton contribution}
S. Schumacher, \emph{et. al.} Phys. Rev. B {\bf 76}, 245324 (2007);
S. Savasta, \emph{et. al.} Phys. Rev. B {\bf 64}, 073306 (2001); S.
Savasta, \emph{et. al.} Semicond. Sci. Technol. {\bf 18}, S294
(2003).


\bibitem{Langbein RRS} R. Houdr\'{e}, R. \emph{et. al.}, Phys. Rev. B {\bf 61},
R13333 (2000); M. Giurioli, \emph{et. al.}, Phys. Rev. B {\bf 64},
165309 (2001); W. Langbein, \& J. M. Hvam, Phys. Rev. Lett. {\bf
88}, 047401 (2002).


\bibitem{Kavokin} K. V. Kavokin, \emph{et. al.}, Phys. Rev. Lett. {\bf 92}, 017401
(2004).

\bibitem{Quattropani+Giacobino} P. Schwendimann, \emph{et. al.}, Phys. Rev. B {\bf 68}, 165324 (2003); J.Ph. Karr,
\emph{et. al.}, Phys. Rev. A {\bf 69}, 063807 (2004).

\bibitem{Piermarocchi PRB} F. Tassone, \emph{et. al.}, Phys. Rev. B {\bf 56}, 7554
(1997).

\bibitem{Lax} M. Lax, Phys. Rev. {\bf 145}, 110 (1966).

\bibitem{notaFDT} Within the Lax approach \cite{Lax}, time dependent noise operators have
quantum statistics microscopically calculated from a nonequilibrium
quantum dissipation-fluctuation theorem with respect to a Markovian
environment. As instance, specialized to our system, we have that
$\langle \mathcal{F}_{P^\dag_{\bf k}}(u) \mathcal{F}_{P_{\bf k}}(v)
\rangle = \delta(u-v) \sum_{\bf k'} W_{{\bf k},{\bf k}'} \langle
\hat{P}^\dag_{\bf k'} \hat{P}_{\bf k'} \rangle (u)$ where $W_{{\bf
k},{\bf k}'}$ are Markovian scattering rates (see Ref.\,
\cite{Portolan PRB 2008a} for further details on polariton systems).


\bibitem{Giacobino} J. Ph. Karr, \emph{et. al.}, Phys. Rev. A {\bf 69},
031802(R) (2004).

\bibitem{EOF} Formally, the EOF is defined as the minimum average pure state
entanglement over all possible pure state decompositions of the
mixed density matrix. Easy speaking the minimum entanglement needed
to construct the density matrix out of some pure states.

\bibitem{Mandel Zinn} See e.g. Zinn-Justin, \textit{Quantum field theory and critical phenomena},
Oxford Science Publications; L. Mandel, \& E. Wolf, \textit{Optical
Coherence and Quantum Optics} (Cambridge University Press) 1995.


\end{thebibliography}
\end{document}